# Automating tumor-infiltrating lymphocyte assessment in breast cancer histopathology images using QuPath: a transparent and accessible machine learning pipeline

Masoud Tafavvoghi [a,*], Lars Ailo Bongo [b], André Berli Delgado [c], Nikita Shvetsov [b], Anders Sildnes [b], Line Moi [d], Lill-Tove Rasmussen Busund [c,d] and Kajsa Møllersen [a]

[a] Department of Community Medicine, Uit The Arctic University of Norway, Tromsø, Norway
[b] Department of Computer Science, Uit The Arctic University of Norway, Tromsø, Norway
[c] Department of Medical Biology, Uit The Arctic University of Norway, Tromsø, Norway
[d] Department of Clinical Pathology, University Hospital of North Norway, Tromsø, Norway

## Abstract

In this study, we built an end-to-end tumor-infiltrating lymphocytes (TILs) assessment pipeline within QuPath, demonstrating the potential of easily accessible tools to perform complex tasks in a fully automatic fashion. First, we trained a pixel classifier to segment tumor, tumor-associated stroma, and other tissue compartments in breast cancer H&E-stained whole-slide images (WSI) to isolate tumor-associated stroma for subsequent analysis. Next, we applied a pre-trained StarDist deep learning model in QuPath for cell detection and used the extracted cell features to train a binary classifier distinguishing TILs from other cells. To evaluate our TILs assessment pipeline, we calculated the TIL density in each WSI and categorized them as low, medium, or high TIL levels. Our pipeline was evaluated against pathologist-assigned TIL scores, achieving a Cohen's kappa of 0.71 on the external test set, corroborating previous research findings. These results confirm that existing software can offer a practical solution for the assessment of TILs in H&E-stained WSIs of breast cancer.

## Introduction

Breast cancer is the most commonly diagnosed cancer among women worldwide, and its incidence is projected to continue rising, with an estimated 1 million deaths annually by 2040 [1]. Histological analysis of tumor-infiltrating lymphocytes (TILs) has become increasingly recognized as a promising biomarker in solid tumors and has reached high-level evidence as a prognostic tool, particularly in human epidermal growth factor receptor 2 (HER2+) and triple negative breast cancer (TNBC) subtypes [2, 3]. However, evaluating TILs through visual methods is still challenging due to a lack of consistent guidelines and adequate training, leading to notable differences in interpretation among observers and introducing variability in results [4, 5]. Therefore, the need for automated image analysis techniques has become essential to reduce human error and variability while ensuring a consistent and precise evaluation of TILs.

Machine learning (ML) algorithms, driven by advancements in computational power and innovative techniques, have significantly transformed the field of computational pathology, providing solutions to many of the limitations associated with the manual assessment of histology slides. Deep learning (DL) models have shown great promise in analyzing whole-slide images (WSI) by learning complex patterns from data. These methods outperform traditional image analysis approaches, which often rely on handcrafted features that may not generalize well across different datasets. These algorithms have shown remarkable performance in many tasks, including tissue and cell segmentation [6, 7], mitosis detection [8], tumor classification [9–11], and prognostication [12, 13]. Applying ML to evaluate TILs in Hematoxylin and Eosin (H&E)-stained histopathological images can reduce observer variability and enhance the reproducibility of assessments.

ML and DL methods for TIL assessment in breast cancer histopathology have rapidly evolved since 2016. Early approaches leveraged traditional image analysis and supervised classifiers. For example, Turkki et al. introduced an "antibody-supervised" deep learning approach by pairing H&E slides with adjacent CD45 immunohistochemistry, using the latter to guide annotations [14]. This method extracted regions (epithelium, stroma, adipose, leukocyte rich) and trained classifiers to quantify immune-cell infiltration, demonstrating that convolutional neural networks (CNNs) could approximate manual TIL counts in H&E-stained slides. Around the same time, researchers began using open-source tools like CellProfiler [15] and Fiji software [16] to detect lymphocyte nuclei via color deconvolution and thresholding. However, such classical techniques yielded only moderate accuracy when compared to ground truth [17]. This motivated a shift toward ML-based detection. QuPath [18] enabled custom ML pipelines–for instance, Bai et al. used QuPath's watershed cell segmentation and a neural network classifier to categorize cells as tumor, TIL, or others [19]. After iterative training and expert review, their classifier achieved >95% accuracy in cell-type labeling. Such integration of human-in-the-loop machine learning in QuPath showed that even pre-deep learning methods could reach high within-cohort performance, though generalizability was limited.

In parallel, fully DL pipelines emerged to tackle TIL assessment end-to-end. A landmark study by Saltz et al. [20] mapped TIL distributions in H&E slides from 13 cancer types in The Cancer Genome Atlas (TCGA) dataset, including breast, using two CNNs–one to detect lymphocyte-rich patches and another to segment necrotic regions. This yielded TIL heatmaps on WSIs, revealing spatial patterns linked to molecular traits. Building on this, Abousamra et al. [21] improved the patch-based TIL classifier with hybrid labels (manual and model-generated) across 23 cancer types. They reported up to 15% higher F1-score than the previous model. Despite these advances, patch-level classification approaches do not explicitly separate stromal from intratumoral TILs, potentially diverging from clinical stromal TIL scoring guidelines [22]. Recent efforts, therefore, focus on the segmentation of tissue compartments followed by fine-grained TIL detection [23] to adhere to international recommendations. Similarly, in the TiGER challenge, top-performing algorithms explicitly performed tumor/stroma segmentation prior to TIL detection [24]. The winning method obtained a tumor–stroma Dice of 0.79 and then localized lymphocytes in stroma, ultimately predicting patient survival with a C-index of 0.719. These results underscore the value of accurate tissue segmentation as a foundation for TIL quantification.

For TIL detection and classification at the cell level, Choi et al. explored an object-detection framework developed using Faster R-CNN to explicitly detect lymphocyte and cancer cell nuclei in WSIs [25]. Their model used a ResNet-34 backbone and optimized a custom Dice-based loss for cell classification. Zhang et al. [26] introduced DDTNet, which improved TIL identification's precision and recall across multiple breast cancer datasets. Another notable toolkit was StarDist, a general DL method for instance segmentation of nuclei based on star-convex shapes. StarDist has been adopted to detect TILs with high precision [27]. This integration of StarDist yielded robust cell detection across variable slide preparations, illustrating how pre-trained models can be repurposed for TIL analysis.

Our study builds on the guidelines for evaluating stromal tumor-infiltrating lymphocyte (sTIL) in breast cancer established by the International sTILs Working Group [28]. These guidelines provide a framework for manual assessment but translating them into ML-based algorithms presents significant challenges. Although some studies have followed these guidelines [25, 29, 30], challenges such as difficulties in accurately segmenting tissue compartments, the lack of guideline-compliant annotated datasets, and the computational demands of fully implementing the recommended workflow have been reported as barriers to their broader adoption [4, 28]. In this study, we developed an end-to-end pipeline solely using QuPath, an open-source, widely used platform to facilitate the integration of ML models into image analysis workflows, providing a transparent and user-friendly approach.

**Methods**

**Datasets description**

In this paper, we used three datasets. For the training of our models, we employed the TIGER_WSIROIS [39] dataset consisting of 195 H&E-stained WSIs of HER2+ and TNBC cases with manually annotated regions of interest (ROIs). These images are sourced from three different collections: 151 from the TCGA-BRCA archive, 26 from Radboud University Medical Center (RUMC), and 18 from the Jules Bordet Institute (JB). Each WSI contains annotated ROIs that mark various tissue compartments, such as invasive and in-situ tumors, tumor-associated stroma, and inflamed stroma. In addition, the annotations of lymphocytes and plasma cells are given as small bounding boxes within the ROIs that can be used to train a cell classifier. Annotations of the TCGA WSIs in this dataset were adapted from the NuCLS [40] and BCSS [41] datasets and relabeled for consistency. Annotations in RUMC and JB subsets were provided by a panel of board-certified breast pathologists. An illustration of tissue and cell level annotations of the TIGER_WSIROIS can be seen in Fig. 1.

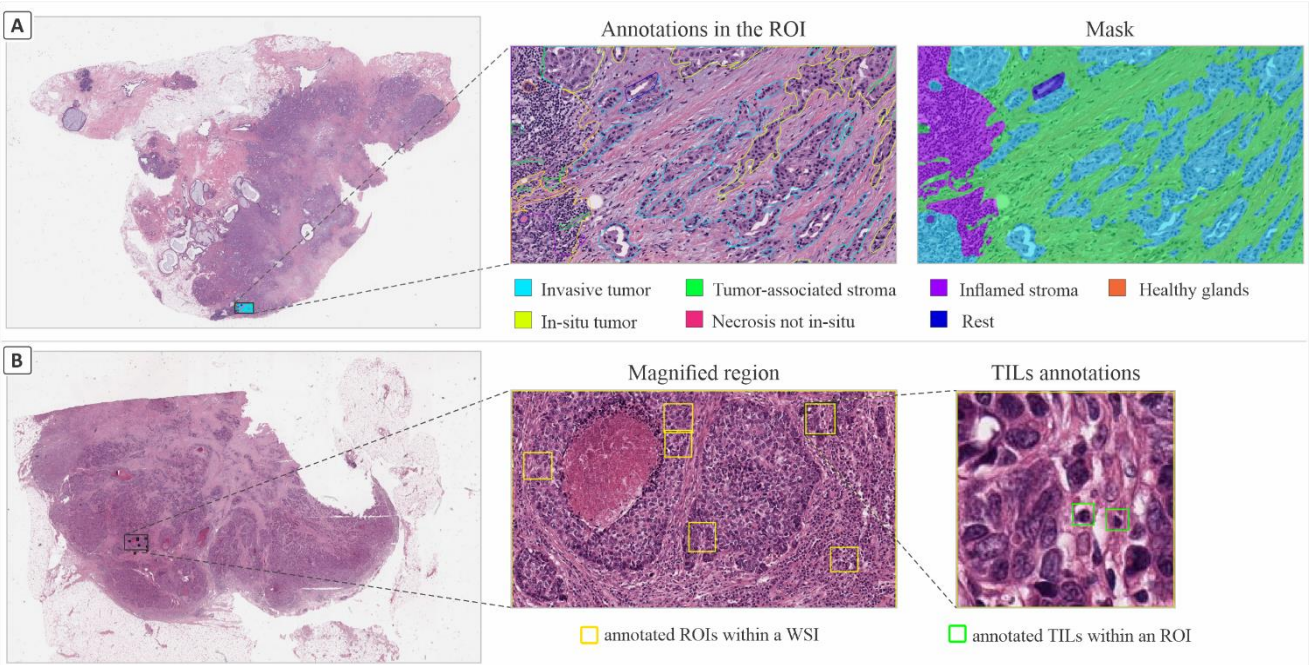

**Figure 1**: Examples of annotations used to train our segmentation and classification models. (A) Annotation of tissue regions in an ROI, where tissue compartments are given in seven classes. (B) Lymphocytes and plasma cells annotations within small ROIs, where all other cell classes are unlabeled in this dataset.

To evaluate our TILs assessment pipeline, we used two other datasets. TIGER_WSITILS [39], a publicly available dataset comprising 82 WSIs of TNBC and HER2+ breast cancer cases from RUMC and JB that were not included in the TIGER_WSIROIS dataset. Each slide in this subset is evaluated visually at the WSI level by a breast pathologist, adhering to the guidelines established by the TILs Working Group. Unlike other datasets, this one does not contain manual annotations; instead, a TIL score for each slide, assessed by a board-certified breast pathologist per the TILs working group's guidelines, is provided as a reference for evaluating our pipeline's performance. The other dataset was an anonymized in-house collection of breast cancer H&E images from the Clinical and Multi-omic (CAMO) cohort [42], including 64 patients diagnosed with HER2+ and TNBC subtypes. Of these, 11 WSIs were excluded based on image quality and the presence and size of diagnostic regions within the tissue, leaving 53 WSIs selected by the pathologist (L.R.B.) as the external test set for our pipeline. An overview of datasets used in this study is shown in Fig. 2.

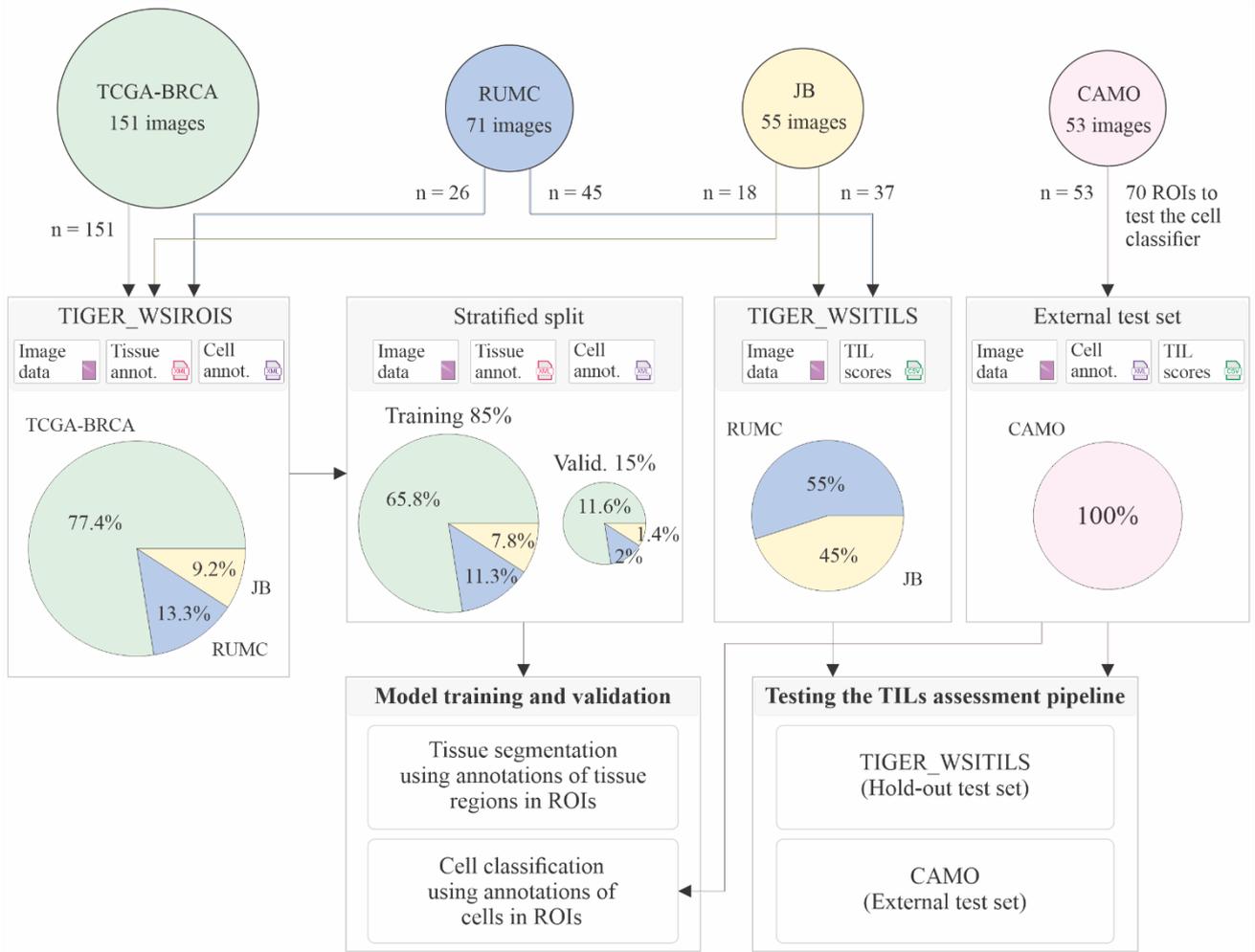

**Figure 2**: An overview of datasets used to train, validate, and test the models in our study.

## Proposed workflow

### Segmentation of tissue compartments

According to the guideline established by the International Immuno-Oncology Biomarker Working Group on Breast Cancer [43], TILs evaluation in H&E-stained slides should focus on the tumor-associated stroma regions of tissue because scoring TILs in the intratumoral compartment is poorly reproducible between pathologists. Therefore, the first step in our workflow was to train a model to segment tissue compartments, with the specific goal of isolating tumor-associated stroma for subsequent analysis.

For the segmentation task, we used the TIGER_WSIROIS dataset and applied a stratified split to divide the data into train (85%) and validation (15%) sets, ensuring adequate representation from TCGA-BRCA, RUMC, and JB in both sets. As shown in Fig. 1, the annotated ROIs were classified into seven categories. Since the TIL score is calculated exclusively in tumor-associated stroma regions, we merged the original classes into three categories: Tumor, which includes invasive and in-situ tumors; Stroma, which encompasses tumor-associated stroma and inflamed stroma; and Other, which combines necrosis, healthy glands, and rest (skin, healthy stroma, adipose tissue, etc.) classes. Based on the description of classes in the TIGER_WSIROIS dataset, the inflamed stroma is part of the tumor-associated stroma characterized by a high concentration of lymphocytes, which was annotated separately by pathologists due to its visual pattern differences.

All WSIs in the TIGER_WSIROIS dataset were published with an approximate spatial resolution of 0.5 $\mu$m/pixel in TIF format, and the annotations were released as XML files. To train a pixel classifier in the QuPath software, we first converted the XML annotations into geojson files using a Python script to make them compatible with QuPath. Then, we used 85% of the annotated ROIs to train a random forest pixel classifier at 1.01 $\mu$m/pixel resolution configured with the following parameters: a maximum tree depth of 30, a minimum sample count of 5, 10 active variables, and a maximum number of 100 trees. The 1SE rule was enabled for model selection. The segmentation model's performance was evaluated using the intersection over union (IoU) and dice coefficient on the 15% validation set of the TIGER_WSIROIS dataset. The workflow of training and evaluation of our tissue segmentation model is shown in Fig. 3.

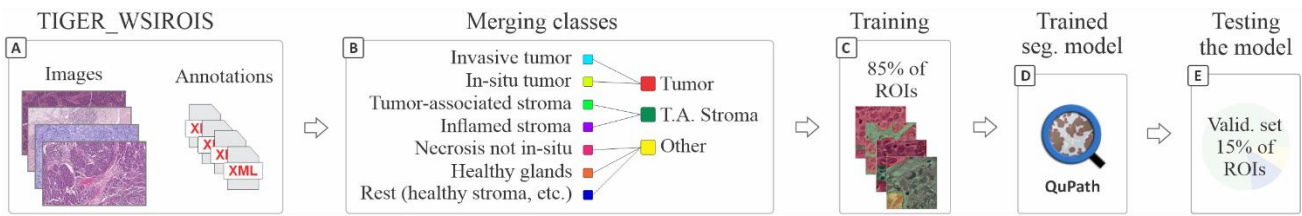

**Figure 3**: The workflow of training a tissue segmentation model. (A) Available image and annotation data. (B) Merging the classes. (C) Training the tissue segmentation model in QuPath. (D) Trained model. (E) Test the trained segmentation model on 15% hold-out validation set.

## Cell classification

For the detection of cells in segmented tumor-associated stroma regions, we used the StarDist [44] plug-in in QuPath with pre-trained weights for H&E images (available at https://github.com/qupath/models/tree/main/stardist). StarDist is a DL-based algorithm that detects and segments nuclei and cells in microscopy images. For 2D images like WSIs, StarDist predicts, for each pixel, the distance to the object's boundary along multiple predefined radial directions, in addition to the likelihood of the object being present. This process generates a comprehensive set of potential object boundaries, which are then refined using non-maximum suppression (NMS) to select the most relevant candidates. The method is particularly effective for detecting spherical structures, such as nuclei and cells, by representing their boundaries as star-convex polygons. Each pixel is associated with a star-convex polygon, determined by the distance from the pixel to the object's boundary along various radial directions, ensuring precise delineation of cells and nuclei and smoothed object features. StarDist was selected for its seamless integration with QuPath, proven accuracy [44, 45], efficiency, and its compatibility with our analysis pipeline.

To train a cell classifier, we employed features of detected cells from the StarDist, together with the TIL annotations provided in the TIGER_WSIROIS dataset. These features, including cells' morphological characteristics and color intensities of channels, in addition to spatial clustering information from the Delaunay triangulation, were all extracted and processed within QuPath. We used 85% of the ROIs with cell annotation for the training and 15% for validating the classifier. Our binary cell classifier was trained in QuPath using a random forest algorithm on a balanced dataset consisting of roughly 34,000 cells. Since only lymphocytes and plasma cells were annotated, we labeled the clearly non-TIL cells as 'Other' to train the model. However, to report the performance of the cell classifier on 15% validation set, we considered all unlabeled cells in the given ROIs as the Other class. The workflow of training our TIL classifier is shown in Fig. 4. In addition to the validation set, we used an external test set, comprising 70 ROIs of size 256×256 pixels with a spatial resolution of 0.5 *μ*m/pixel from the 53 selected WSIs in the CAMO cohort. All TILs within these ROIs were annotated by the pathologist (L.R.B.) using QuPath software.

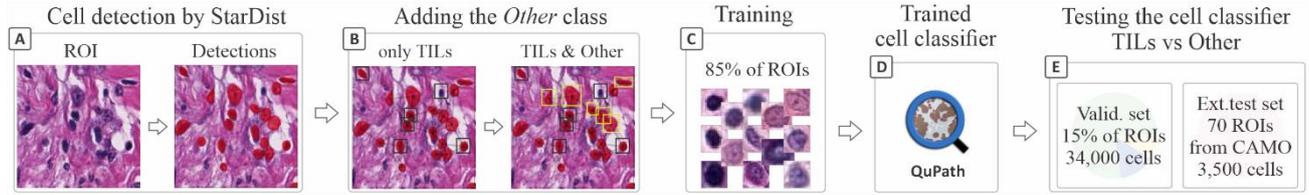

**Figure 4**: The workflow of training the cell classifier in QuPath. (A) Detection and segmentation of cells in QuPath in the given ROIs using the pre-trained StarDist algorithm. (B) Annotating non-TIL cells to create a new class (Other) for training a binary cell classifier. (C) Training the cell classifier in QuPath. (D) Trained cell classifier to count TILs in WSIs. (E) testing the trained model on 15% validation set and external test set.

### Evaluating the pipeline

To evaluate the whole pipeline, the pathologist-provided TIL scores in the TIGER_WSITILS dataset served as the reference for the evaluation. This subset of data has 82 WSIs from the RUMC and JB collections that were not used in any part of our model developments. We refer to this as the hold-out test set in this paper. To process the WSIs, first, we applied the trained tissue segmentation model to separate tumor-associated stroma from other compartments in the tissue. Then, the StarDist cell detection algorithm was applied to the segmented stroma region for the subsequent TILs classification and quantification (Fig. 5). Unlike ground truth TIL scores for WSIs in the TIGER_WSITILS test set that were given as a percentage, our pipeline measures TILs density as the number of TILs per millimeter square of tumor-associated stroma. Thus, to harmonize the evaluations for comparison, we categorized both pathologist scores and TILs densities from our pipeline into Low, Medium, and High levels. TILs scores were categorized as follows: Low (<10%), Medium (10–40%), and High (>40%) according to the International sTILs Working Group recommendation; and TILs density values from the pipeline were stratified into the same levels (Low, Medium, High) based on their percentage relative to the maximum observed TILs density. Additionally, to evaluate our TILs assessment pipeline on an external test set, we used 53 WSIs from the CAMO cohort. For each WSI, the pathologist (L.R.B.) provided a reference TILs score level, which was used for comparison with the pipeline predictions based on the same criteria.

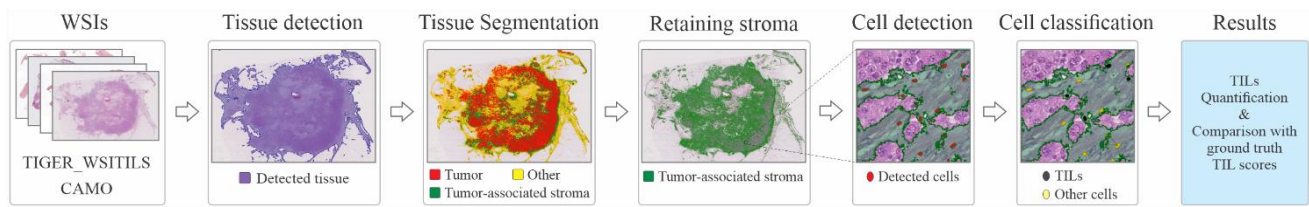

**Figure 5**: The workflow of testing the TILs assessment pipeline in QuPath.

### Results

### Segmentation of tissue compartments

We used IoU and dice coefficients to evaluate the tissue segmentation model's performance metrics. The mean IoU and mean Dice scores for the validation set are summarized in Table 1 for each class. The segmentation model demonstrated moderate performance, with a mean IoU of 0.62 and a mean dice score of 0.70. The model demonstrated a balanced performance in segmenting both the tumor and tumor-associated stroma while achieving relatively higher accuracy in detecting the Other class. These results suggest that the model is effective in segmenting the tissue regions, though further improvements could be explored to enhance tumor segmentation. Examples of segmentation model predictions are illustrated in Figure 6, showing the model's adequate ability to delineate different tissue compartments. The segmented regions closely match the ground truth annotations, highlighting the ability of the model to identify different tissue compartments.

**Table 1**: Performance of the segmentation model on the validation set.

| mIoU_Tum. | mIoU_Str. | mIoU_Other | overall IoU | mDice_Tum. | mDice_Str. | mDice_Other | overall Dice |
|---|---|---|---|---|---|---|---|
| 0.558 | 0.597 | 0.701 | 0.616 | 0.673 | 0.731 | 0.796 | 0.703 |

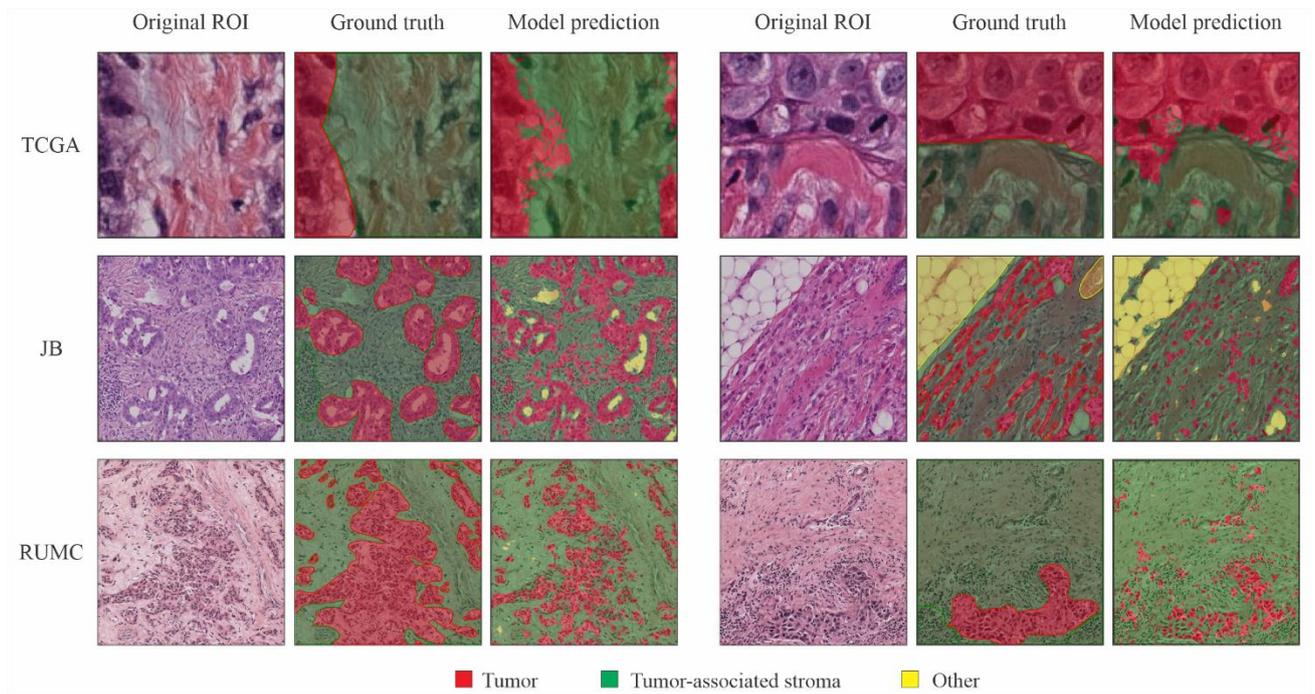

**Figure 6**: Examples of segmentation model predictions for six ROIs from WSIs in the TCGA-BRCA, JB, and RUMC subsets in the validation set.

**Cell classification**

The performance of the cell classification model was evaluated on two distinct sets: a 15% validation set and an external test set from the CAMO cohort. On the validation set, the model achieved an accuracy of 0.81, with a weighted average F1-score of 0.83. The area under the receiver operating characteristic curve (AUC-ROC) was 0.865, with a 95% confidence interval of (0.859, 0.871), indicating good classification performance. Precision for TILs was 0.44, with recall at 0.89, reflecting that the model identified TILs with high sensitivity but at the cost of a relatively higher number of false positives, as seen in the confusion matrix with 460 false positives and 3,735 true positives (Figure 7A).

In contrast, the model performed better on the external test set, achieving an accuracy of 0.96, with a weighted average F1-score of 0.96. The AUC-ROC was 0.944, with a 95% confidence interval of (0.929, 0.956), demonstrating excellent discrimination between TILs and other cells. Precision and recall for TILs were 0.81 and 0.91, respectively, suggesting that the model maintained a good balance between correctly identifying TILs and minimizing false positives, as shown by the confusion matrix with 99 false positives and 426 true positives (Figure 7B).

These results highlight the model's strong performance in distinguishing TILs from other cells across different datasets. The model achieved superior accuracy and precision on the external test set compared to the validation set, an unexpected outcome that is further explored in the Discussion section. More details of the cell classifier's performance on validation and test sets are summarized in Table 2, and an illustration of cell classification is shown in Figure 7D.

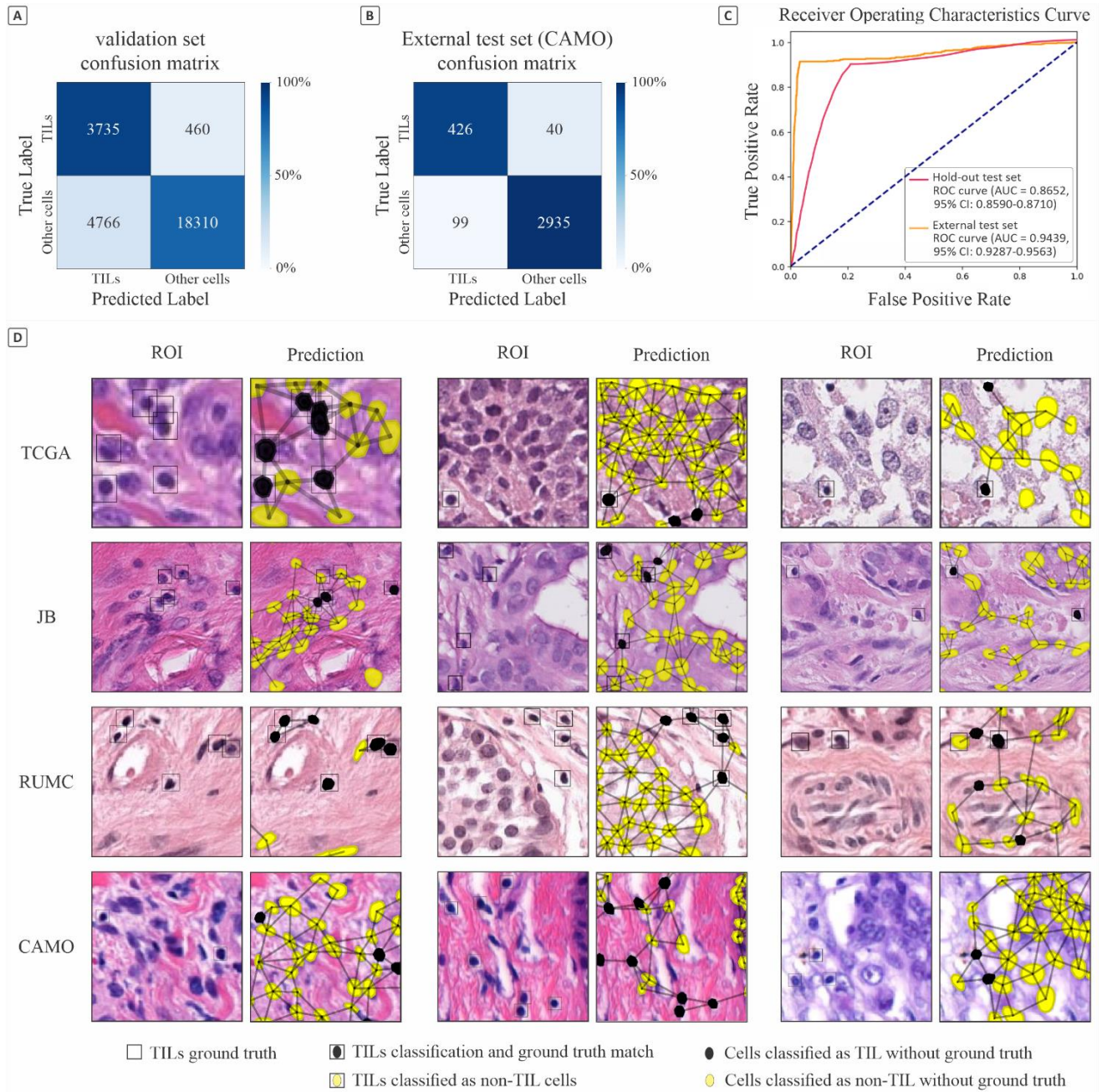

**Figure 7**: Cell classifier's performance on the validation and test set ROIs. (A) Confusion matrix of the validation set. (B) Confusion matrix of the external test set. (C) ROC curves of classification of TILs versus other cells in validation and external test sets. (D) An illustration of the cell classifier's predictions, where black squares show pathologists' annotated TILs. Black and yellow overlays represent cells classified as TILs or other (non-TIL) cells, respectively. The lines between the cells result from Delaunay triangulation, which is used to represent the spatial relationships between cells for subsequent classification. The difference between cell sizes in the figure is due to different ROI sizes in the dataset.

**Table 2**: Performance of the cell classification model on validation and external test sets.

|  | Hold-out validation set | | | External test set (CAMO) | | |
|---|---|---|---|---|---|---|
| Class | Precision | Recall | F1-score | Precision | Recall | F1-score |
| TILs | 0.44 | 0.89 | 0.59 | 0.81 | 0.91 | 0.86 |
| Other | 0.98 | 0.79 | 0.88 | 0.99 | 0.97 | 0.98 |
| Weighted avg. | 0.89 | 0.81 | 0.83 | 0.96 | 0.96 | 0.96 |

**TILs score evaluation**

To evaluate the whole TILs assessment pipeline, we used TIGER_WSITILS and the CAMO datasets, where the TILs scores were categorized into low, medium, and high levels. For the hold-out test set (TIGER_WSITILS), Cohen's Kappa value between the pathologist and our pipeline was 0.685, indicating moderate agreement. The confusion matrix in Figure 8 shows that the model has high precision and recall for the medium and low categories, with the high category exhibiting slightly lower performance (precision of 0.75). Overall, our pipeline achieved an accuracy of 0.79 with a weighted F1-score of 0.79, which reflects balanced performance across the categories, relative to pathologists' TIL assessments as the reference.

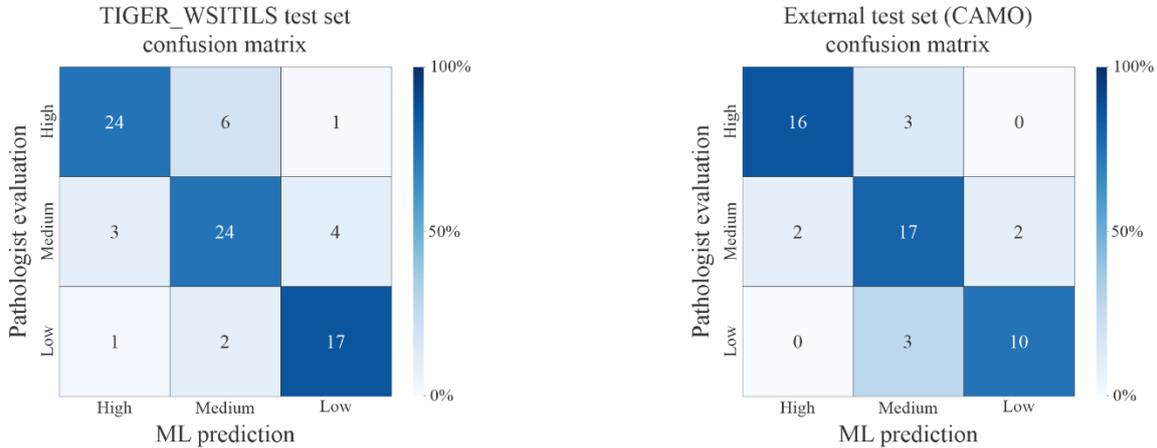

**Figure 8**: Performance of our pipeline in classification of TILs score levels compared to pathologist evaluations.

For the CAMO test set, our pipeline showed a Cohen's Kappa value of 0.710, indicating similar agreement with the pathologist evaluation of TILs levels as observed in the hold-out test set. The confusion matrix of the TILs score classification (Figure 8) indicates that the model has very high precision (0.89) and recall (0.84) for the medium category, while the high category again showed slightly lower precision (0.74). The overall accuracy for this dataset was 0.82, with a weighted average F1-score of 0.81, further emphasizing the pipeline's consistent performance across all categories (Table 3).

**Table 3**: Comparison of TILs assessment scores from our pipeline and pathologists' evaluations for the test sets.

|  | TIGER_WSITILS dataset | | | External test set (CAMO) | | |
|---|---|---|---|---|---|---|
| Level of TILs | Precision | Recall | F1-score | Precision | Recall | F1-score |
| Low | 0.77 | 0.85 | 0.81 | 0.83 | 0.77 | 0.80 |
| Medium | 0.86 | 0.77 | 0.81 | 0.89 | 0.84 | 0.86 |
| High | 0.75 | 0.75 | 0.76 | 0.74 | 0.81 | 0.77 |
| Weighted avg. | 0.80 | 0.79 | 0.79 | 0.82 | 0.81 | 0.81 |

## Discussion

In this study, we developed a pipeline for the assessment of stromal TILs in H&E-stained WSIs of breast cancer. The pipeline consisted of three main components, all implemented within the QuPath software: (1) segmentation of tissue compartments to isolate tumor-associated stroma from other regions, (2) detection and classification of cells within the tumor-associated stroma, and (3) quantification of TILs in this region.

The segmentation model successfully delineates tumor-associated stroma, which is essential for accurate TILs assessment, achieving a mean Dice score of 0.73 for stroma segmentation (Table 1). This demonstrates the model's capability in identifying tumor-associated stroma regions. The segmentation of the Other class, including regions such as necrosis and healthy tissue, showed particularly a better performance, likely due to the more distinct boundaries of these regions compared to tumor and stroma. While the model performed well in segmenting tumor-associated stroma, tumor segmentation could be further improved. Nevertheless, both the tumor and Other regions were excluded from subsequent TILs detection and quantification.

Compared with other studies, the segmentation accuracy of our model falls within the mid-range. For example, HookNet-based segmenters in the TIGER challenge achieved a Dice score of 0.72 on external tests, while the MuTILs panoptic model reported a higher Dice score of 0.81 for tumor-associated stroma segmentation, likely benefiting from extensive training on over 16,000 annotated regions. Shephard et al. [24] reported a Dice score of 0.79 for tumor–stroma segmentation using a dual CNN approach. While larger datasets and more complex architectures can yield higher accuracy, methods using smaller datasets or single-institution data tend to report Dice scores in the 0.6–0.75 range, similar to our method.

Nevertheless, our decision to implement the pipeline fully within the QuPath software, a widely used tool in the pathology community, was driven by the goal of providing an accessible solution for researchers and pathologists. Although more complex methods could yield better segmentation results, using already available ML algorithms and extensions within QuPath allowed us to maintain simplicity, ease of use, and seamless integration, ensuring a flexible and user-friendly pipeline.

Our cell classification results highlighted the model's ability to generalize well to unseen data. However, the performance on the validation set was notably lower, which warrants further investigation. We hypothesize that the discrepancy in performance between these sets could be attributed to errors in the annotation of TILs in the validation set. Specifically, for the validation set, the TILs class achieved a high accuracy of 0.89, but the low precision of 0.44 indicates that false positives may have played a significant role. This suggests that many cells were classified as TILs, while in the ground truth, they were considered as Other cells. One potential explanation of this is the class imbalance, where TILs and Other cells had a 15% to 85% distribution. Thus, a 20% misclassification in the major class can significantly impact the precision for the less frequent class. The other possible cause of low precision in the validation set is that some TILs within the ROIs were not annotated (Figures 9), leading to a mismatch between the ground truths and the model's predictions for those cells. In addition, there were obvious errors, such as misclassified tumor cells as TILs in the provided labels, which could further contribute to false negatives and reduce the classification recall (sensitivity) (Figures 9). These annotation inconsistencies, typical in histological image analysis, likely caused the model's moderate precision for TILs class on the validation set.

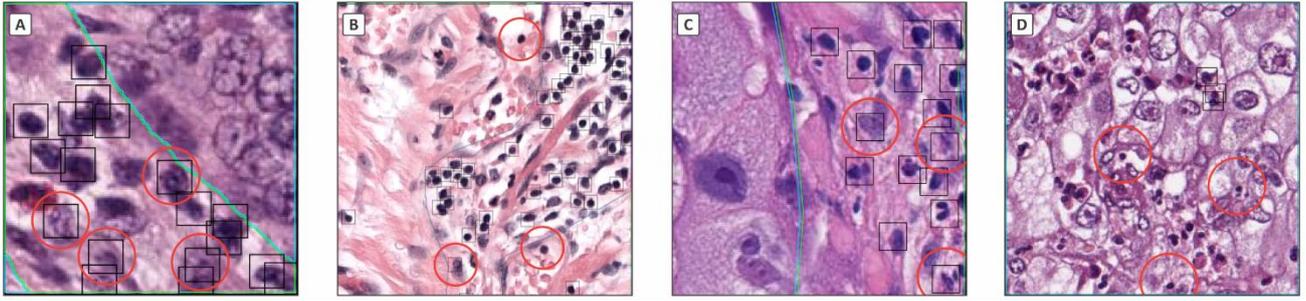

**Figure 9**: Examples of errors in the annotations of cells in the TIGER_WSIROIS dataset. Black boxes show cells annotated as TILs, and red circles indicate possible incorrect annotations. The difference in cell sizes in the figure is due to varying ROI sizes in the dataset. (A) Examples of mislabeling, where non-TIL cells are annotated as TILs. (B) Unlabelled TILs in an ROI and a possible mislabeling. (C) Non-TIL cells annotated as TILs. (D) Possible unlabeled TILs that can lead to more false positive predictions by the ML model.

The improved performance on the smaller external test set (CAMO cohort), which contained fewer cells (3,500 compared to 27,271 in the validation set), may be attributed to more accurate and consistent annotations. With a smaller dataset, the impact of annotation errors, such as misclassified or missed TILs, was likely reduced, allowing for more reliable cell classification. The smaller sample size may have also facilitated a more manageable and precise annotation process, contributing to the model's better performance, achieving a precision of 0.81. These findings underscore the critical role of high-quality, consistent annotations in training and evaluating ML models, particularly in medical imaging, where variations in annotation quality and dataset characteristics can significantly influence model performance and lead to biased or misleading results.

In comparison, the cell classification performance of our model aligns well with similar studies in the field, although some have used a TILs detection approach, unlike our cell classification method. For instance, Shephard et al. [24] reported an F1-score of 0.702 for TILs detection, which is lower than the 0.83 weighted average F1-score achieved by our cell classification method. The MuTILs model also attained an AUC-ROC of 0.93 for lymphocyte detection using cross-validation [37], which is comparable to our study's AUC-ROCs for the validation and test sets. Additionally, a study by Makhlouf et al. [46] reported an average F1-score of 0.82 for immune cell classification, where they categorized cells into tumor, immune, and stroma classes.

Pathologist-provided TILs scores are inherently subjective, with variability in how different evaluators interpret and assess TILs within tissue regions. Such inconsistencies are a common challenge in histological image analysis. Accurately defining boundaries between tissue compartments in breast cancer slides is also difficult due to tissue heterogeneity, making it challenging to establish clear delineations between tumor-associated stroma and surrounding areas. These factors contribute to the observed discrepancies and highlight the importance of reliable, consistent annotations for robust model evaluation.

The TILs score, as defined by the International sTILs Working Group, calculates the area of tumor-associated stroma occupied by TILs relative to the total tumor-associated stroma area. While this approach is useful in clinical settings, it may not translate well to the quantitative approach used in our pipeline, where TILs density (the number of TILs per square millimeter of tumor-associated stroma) is calculated. Consequently, both the pathologist's percentage scores, and the pipeline's density measurements were categorized into three levels—low, medium, and high—for a fair comparison. While the ML pipeline calculates TILs density, it is not without challenges, including potential over- or under-counting of cells in densely populated or sparse regions. However, it avoids some of the subjective nature inherent in percentage-based scoring, where small differences in boundary definitions can lead to significant variations in reported TIL percentages.

In terms of agreement with pathologist-provided scores, our pipeline achieved a Cohen's kappa of 0.685 and 0.710, indicating a moderate to substantial agreement. This is comparable to other studies. For example, the TILs scores computed by the MuTILs model showed a moderate correlation with visual (pathologist) scores, with Spearman correlations ranging from 0.55 to 0.61. A recent study by Choi et al. [25] achieved a concordance correlation coefficient of 0.755 between the DL model-derived sTIL scores and the average of pathologists. The study found substantial variation in sTIL scores provided by different pathologists, with almost half of the cases showing a greater than 10% difference in their evaluations. Two other studies [24, 46] reported similar agreement between the visual and AI-based TIL scores, with a Pearson correlation coefficient of 0.744 and an intraclass correlation coefficient of 0.7, respectively.

It is important to note that some variability in ground truth exists even among experts. Pathologist assessments of the same cases often differ, with inter-observer kappa values typically ranging between 0.57 and 0.78 [47–50]. Therefore, a kappa of around 0.7 for automated versus human assessment indicates agreement on par with inter-pathologist concordance. Regarding prognostic stratification, the high agreement with expert assessments suggests that the model can reliably capture prognostic signals, such as distinguishing between high and low TIL cases. This demonstrates that our model's performance aligns closely with expert evaluations, supporting its potential for reliable application. While performance metrics across studies are not always directly comparable due to different datasets and evaluation protocols, our TILs assessment pipeline achieves results that are on par with other DL-based methods. Its adherence to stromal TIL density aligns with clinical guidelines and is likely to provide a similarly robust prognostic indicator, as studies have consistently found that stromal TIL scores are significantly associated with survival. The integration of this method into QuPath further enhances its usability, allowing for seamless application to WSIs and pathologist oversight, addressing practical gaps that have only recently been tackled in the literature.

## Conclusion

In this study, we developed an end-to-end, accessible ML pipeline solely within the QuPath platform for the automatic assessment of TILs in H&E-stained WSIs of breast cancer. Our approach demonstrated the potential of ML models to provide a practical solution for TIL evaluation, with performance comparable to that of pathologists and other reported studies.